\documentstyle[sprocl]{article}
\begin{document}
\bibliographystyle{unsrt}    

\newcommand{\st}{\scriptstyle}
\newcommand{\sst}{\scriptscriptstyle}
\newcommand{\mco}{\multicolumn}
\newcommand{\epp}{\epsilon^{\prime}}
\newcommand{\vep}{\varepsilon}
\newcommand{\ra}{\rightarrow}
\newcommand{\ppg}{\pi^+\pi^-\gamma}
\newcommand{\vp}{{\bf p}}
\newcommand{\ko}{K^0}
\newcommand{\kb}{\bar{K^0}}
\newcommand{\al}{\alpha}
\newcommand{\ab}{\bar{\alpha}}
\def\be{\begin{equation}}
\def\ee{\end{equation}}
\def\bea{\begin{eqnarray}}
\def\eea{\end{eqnarray}}
\def\CPbar{\hbox{{\rm CP}\hskip-1.80em{/}}}

\def\alt{\stackrel{<}{\sim}}
\def\agt{\stackrel{>}{\sim}}
\def\eslt{E\llap/_T}
\def\etmiss{E\llap/_T}
\def\tg{\tilde g}
\def\tt{\tilde t}
\def\tq{\tilde q}
\def\tf{\tilde f}
\def\tl{\tilde \ell}
\def\tell{\tilde \ell}
\def\tnu{\tilde \nu}
\def\te{\tilde e}
\def\tmu{\tilde \mu}
\def\ttau{\tilde \tau}
\def\tz{\widetilde Z}
\def\tw{\widetilde W}

\def\ap#1#2#3   {{\em Ann. Phys. (NY)} {\bf#1} (#2) #3}
\def\apj#1#2#3  {{\em Astrophys. J.} {\bf#1} (#2) #3}
\def\apjl#1#2#3 {{\em Astrophys. J. Lett.} {\bf#1} (#2) #3}
\def\app#1#2#3  {{\em Acta. Phys. Pol.} {\bf#1} (#2) #3.}
\def\ar#1#2#3   {{\em Ann. Rev. Nucl. Part. Sci.} {\bf#1} (#2) #3}
\def\cpc#1#2#3  {{\em Computer Phys. Comm.} {\bf#1} (#2) #3}
\def\err#1#2#3  {{\it Erratum} {\bf#1} (#2) #3}
\def\ib#1#2#3   {{\it ibid.} {\bf#1} (#2) #3}
\def\jmp#1#2#3  {{\em J. Math. Phys.} {\bf#1} (#2) #3}
\def\ijmp#1#2#3 {{\em Int. J. Mod. Phys.} {\bf#1} (#2) #3}
\def\jetp#1#2#3 {{\em JETP Lett.} {\bf#1} (#2) #3}
\def\jpg#1#2#3  {{\em J. Phys. G.} {\bf#1} (#2) #3}
\def\mpl#1#2#3  {{\em Mod. Phys. Lett.} {\bf#1} (#2) #3}
\def\nat#1#2#3  {{\em Nature (London)} {\bf#1} (#2) #3}
\def\nc#1#2#3   {{\em Nuovo Cim.} {\bf#1} (#2) #3}
\def\nim#1#2#3  {{\em Nucl. Instr. Meth.} {\bf#1} (#2) #3}
\def\np#1#2#3   {{\em Nucl. Phys.} {\bf#1} (#2) #3}
\def\pcps#1#2#3 {{\em Proc. Cam. Phil. Soc.} {\bf#1} (#2) #3}
\def\pl#1#2#3   {{\em Phys. Lett.} {\bf#1} (#2) #3}
\def\prep#1#2#3 {{\em Phys. Rep.} {\bf#1} (#2) #3}
\def\prev#1#2#3 {{\em Phys. Rev.} {\bf#1} (#2) #3}
\def\prl#1#2#3  {{\em Phys. Rev. Lett.} {\bf#1} (#2) #3}
\def\prs#1#2#3  {{\em Proc. Roy. Soc.} {\bf#1} (#2) #3}
\def\ptp#1#2#3  {{\em Prog. Th. Phys.} {\bf#1} (#2) #3}
\def\ps#1#2#3   {{\em Physica Scripta} {\bf#1} (#2) #3}
\def\rmp#1#2#3  {{\em Rev. Mod. Phys.} {\bf#1} (#2) #3}
\def\rpp#1#2#3  {{\em Rep. Prog. Phys.} {\bf#1} (#2) #3}
\def\sjnp#1#2#3 {{\em Sov. J. Nucl. Phys.} {\bf#1} (#2) #3}
\def\spj#1#2#3  {{\em Sov. Phys. JEPT} {\bf#1} (#2) #3}
\def\spu#1#2#3  {{\em Sov. Phys.-Usp.} {\bf#1} (#2) #3}
\def\zp#1#2#3   {{\em Zeit. Phys.} {\bf#1} (#2) #3}

\setcounter{secnumdepth}{2} 

\hfill{UH-511-834-95}
\vspace{5mm}
\title{SUPERSYMMETRY PHENOMENOLOGY: A MICROREVIEW~\protect\footnote{Presented
at
the 1995 European Physical Society Meeting, Brussels, Begium, July 1995.}}

\author{XERXES TATA}

\address{Department of Physics and Astronomy, University of Hawaii,
Honolulu, HI 96822, USA}
\maketitle
\abstracts{We briefly review the current status and future prospects
for supersymmetry searches at colliders, and discuss strategies
by which further information about sparticle properties may be obtained
at the LHC.}
\section{Current Status of Supersymmetry}

{\it Direct Constraints from Colliders:}
There is no evidence for the production of sparticles
in high energy collisions. Direct searches~\cite{PDG} for sparticles
in experiments at LEP yield lower bounds $\sim 45$~GeV
on the masses of charginos ($\tw_1$), charged sleptons
and squarks. While these bounds are sensitive to how sparticles
decay, the corresponding limits~\cite{BDT} (which also apply to $m_{\tnu}$)
from the $Z$ line shape~\cite{OLCH} ({\it i.e.} from the
measured values of $\Gamma_Z$ and $\Gamma_Z^{inv}$) which
are only slightly weaker, are insensitive to
specific sparticle decay patterns.
Furthermore, since virtual effects of sparticles rapidly
decouple as $m_{SUSY} \rightarrow \infty$,
the agreement~\cite{OLCH} of precision measurements at LEP with Standard Model
(SM) expectation can be readily accommodated if
sparticles are heavier than about 200~GeV: by the same
token, if say chargino-top squark
loop effects are assumed to be responsible for the deviation
in $R_b$, these should probably be discoverable~\cite{KANE}
at LEP2, or from an analysis~\cite{SENDER}
of the data from the current Tevatron run. The quoted value of $R_c$
is more difficult~\cite{SOLA} to accommodate.
The best limits on
gluinos and squarks come from experiments at hadron colliders.
The non-observation of an excess of $\eslt$ events has enabled
the D0 and CDF experiments to infer~\cite{NODUL} a limit of 173~GeV (229~GeV)
on the mass of the gluino if squarks are very heavy (for $m_{\tq}=m_{\tg}$).
We should not be discouraged by the absence of SUSY signals
in current experiments which have probed masses up
to 50~GeV (200~GeV) for weakly (strongly) interacting sparticles. In
comparison,
the natural mass scale for sparticles is 100-1000~GeV
in order for supersymmetry to be able to stabilize the elementary
scalar sector thought to be responsible for electroweak symmetry
breaking.

{\it Framework for SUSY Analysis:}
The minimal supersymmetric model (MSSM) with no other assumptions
other than the particle content, the low energy symmetry group,
and $R$-parity conservation, leads to a plethora of new parameters
(especially in the soft-SUSY breaking sector where we parametrize
our ignorance about the physics of SUSY breaking),
making phenomenological
analyses intractable. For this reason, most analyses today are
done within the minimal supergravity (SUGRA) GUT framework~\cite{DREES}
with the radiative
breaking of electroweak symmetry. The assumed symmetries about
the physics at the high scale then ensure that all sparticle
masses and couplings are fixed by just four parameters:
a common soft SUSY breaking mass $m_0$ for all scalars, a common SUSY breaking
gaugino mass $m_{1/2}$, a common value ($A_0$) for trilinear scalar couplings
and $\tan\beta$, the ratio of the VEVs of the two Higgs fields
in the model. There is some ambiguity about exactly what the scale ($M_X$),
at which the running scalar masses and $A$-parameters unify, is. We
take $M_X$ to be the scale $M_{GUT}$ at which the gauge couplings
unify. If $M_X>M_{GUT}$, as it might well be, the scalar masses
and $A$-parameters would not be exactly universal at $M_X$. The
biggest impact of this would probably be on the condition
of radiative symmetry breaking~\cite{POK} which fixes the superpotential
Higgsino mass ($\mu$) squared (notice that the sign of
$\mu$ is not fixed). Allowing $\mu$ to be
a free parameter would be tantamount to relaxing this latter
constraint. We stress that the assumptions underlying the
SUGRA GUT framework may ultimately prove to be incorrect. It
is, therefore, important to devise ways by which these might
be tested in future experiments, as well as to examine
the sensitivity of various signals to these assumptions,
particularly when trying to determine the capabilities
of future facilities.

\section{Sparticle Searches at Future Colliders}

The LEP collider will soon enter its second phase and its energy
will ultimately be upgraded to 175-200~GeV. Given a data sample
of $\sim 300 \ pb^{-1}$, the clean environment of $e^+e^-$ collisions
should make it possible to search~\cite{LEP2} for charginos, sleptons, squarks,
and even Higgs bosons, up to 80-90~GeV ($b$-tagging capability may
be needed if $m_H \simeq M_Z$). The corresponding mass reach
for neutralinos is sensitive to their (parameter-dependent) mixing
angles. An analysis within the SUGRA framework, where various cross
sections are correlated, has also been performed~\cite{BRHLIK}.

The CDF and D0 experiments have each already accumulated
an integrated luminosity of about 100~$pb^{-1}$. With a sample
of this size, they will be able to substantially
extend~\cite{KAMON,MRENNA} their search for gluinos and squarks in the $\eslt$
channel. Moreover, they should also be able to begin searching for
multilepton events~\cite{RPV} from their cascade decays:
the most promising of which are the same-sign (SS) isolated
dileptons plus jets  plus $\eslt$ and the isolated trileptons plus jets
plus $\eslt$ events. The SM {\it physics}
backgrounds in these channels are tiny
so that a conclusive observation of even a handful of events in these
channels could signal new physics. Although the cross sections for these
spectacular signatures is small, these rate-limited channels may already
begin to
be competetive with the background-limited $\eslt$ channel with a reach
around 200~GeV (250-300~GeV) if squarks are heavy (if $m_{\tq}=m_{\tg}$)
by the time this data is analysed. At the Main Injector (MI), where we expect
an order of magnitude larger data sample, SUSY searches via these
multilepton channels will be very important.

Tevatron experiments
should also be able to search~\cite{AN} for the direct production
of charginos and neutralinos ($\tz_i$). The most promising channel is
$p\bar{p} \to \tw_1\tz_2 \to \ell\nu\tz_1+\ell'\bar{\ell'}\tz_1$,
which leads to spectacular events with three, hard isolated leptons
and $\eslt$ and essentially no jet activity. In fact, the preliminary
analyses~\cite{NODUL} of the Run IA data already yield lower
bound on $m_{\tw_1}$ competitive with those from LEP.
Assuming the unification of gaugino masses, Tevatron experiments
should be able to (indirectly) probe~\cite{TEVSTAR,MRENNA}
gluino masses up to 400-500~GeV
(550-700~GeV) at the MI (TeV33) upgrades for favourable ranges
of parameters. Confirmatory signals from chargino pair
production may also be present~\cite{TEVSTAR}.
There are, however, regions of parameter space
where the branching fraction for the leptonic decay of $\tz_2$
is strongly suppressed so that there is no observable $3\ell$
signal~\cite{TEVSTAR,MRENNA}
even if the chargino is at its current bound from LEP.

Experiments at the Tevatron should also be able to search for
the less massive of the two
top squarks ($\tt_1$) which, because of its large Yukawa interaction,
can be substantially lighter than all other squarks even within
the SUGRA framework. If $\tt_1$ is heavy enough it will dominantly
decay via $b\tw_1$: since the chargino decays via $\tw_1 \to f\bar{f'}\tz_1$,
signals from $\tt_1\bar{\tt_1}$ production will then be identical to those
from top quark pair production. If the chargino decay mode of $\tt_1$
is kinematically inaccessible, it decays~\cite{HK} via $\tt_1 \to c\tz_1$ so
that top squark pair production is then signalled by multi-jet plus
$\eslt$ events. It has been shown~\cite{SENDER}
that, irrespective of how $\tt_1$
decays, with an integrated luminosity of $\sim 100 \ pb^{-1}$, experiments
at the Tevatron should be able to find it if $m_{\tt_1}\alt 100$~GeV (some
$b$-tagging capability is needed for the detection of a leptonically
decaying top squark in the 1$\ell+jets+\eslt$ channel). Indeed
a preliminary analysis by the D0 Collaboration~\cite{WIGHT} excludes
60~GeV$\alt m_{\tt_1}\alt 100$~GeV if $m_{\tz_1}\alt 30$~GeV,
and $\tt_1$ decays via $\tt_1\to c\tz_1$.
A
recent analysis~\cite{MRENNA} suggests that MI experiments should be able to
detect a $t$-squark up to 160~GeV if it decays via the chargino mode.
Finally, it has been shown~\cite{SLEP}
that it difficult to search for sleptons much heavier
than 50~GeV even at the MI upgrade of the Tevatron.

Gluino and squark
searches at the LHC are discussed by Polesello~\cite{POL}, so we will
limit ourselves to simply stating that experiments there should be
able to detect~\cite{ATLAS} gluinos as heavy as 1300~GeV (2~TeV) if the squark
is heavy ($m_{\tq}=m_{\tg}$) in the $\eslt$ channel. A reach of
about 1~TeV should also be possible~\cite{BTW,ATLAS,DP}
via the SS and multilepton channels. The $\eslt$~\cite{ETM} and
leptonic~\cite{PREP} signals have been studied within the SUGRA
framework: the largest reach is via the $1\ell$ channel.
It also appears that there is no window of masses where gluinos
will evade detection both at the LHC and at the MI.
What about other sparticles?
LHC experiments should be able to discover sleptons~\cite{AMET,SLEP} with
masses up to about 250~GeV in the $\ell^+\ell^- +\eslt$ channel,
provided jets can be vetoed
with high efficency to eliminate backgrounds
from $t\bar{t}$ production. It should also be possible
to search for trileptons~\cite{TRILEP} from
$\tw_1\tz_2$ production except in those parameter ranges
where the leptonic decay of $\tz_2$ are strongly suppressed.
This signal sharply cuts off if $\tz_2$ is heavy enough
to decay into real $Z$ or Higgs
bosons. The reach of the Tevatron and its upgrades and the LHC
are compared in a recent phenomenological review~\cite{DPF}.

Discovering
charged sparticles (also Higgs bosons)
with masses essentially all the way to the
beam energy is easy~\cite{FUJII} at a linear $e^+e^-$ colliders, provided
sufficient integrated luminosity is attained (10-30~$fb^{-1}$ at
$\sqrt{s}=500$~GeV). We stress here that the experiments at Tevatron
upgrades (including TeV33), while very interesting, cannot explore the
complete parameter space of weak scale SUSY. For this, the LHC
or a Linear Collider with $\sqrt{s}=500$-1000~GeV are essential.

\section{Beyond Sparticle Searches}

Fujii~\cite{FUJII} has described how the precision measurements of sparticle
masses that can
be made with polarized beams at Linear Colliders can be used
to incisively test the assumptions underlying the SUGRA GUT framework.
At these machines, one would also be able to check~\cite{FENG}
(to within $\sim 30$\%) whether sparticle couplings
are related to the known particle couplings as predicted by SUSY. Here, we will
briefly discuss what other interesting things LHC experiments might
be able to do if SUSY is discovered.

{\it Mass Measurements:} For a wide range of parameters, it is
possible to isolate $\ell^+\ell^-\ell'$
events from $\tw_1\tz_2$ production from SM as well as other SUSY
sources: thus, the end point of the $m(\ell^+\ell^-)$ distribution
yields~\cite{TRILEP}
a reliable measure of $m_{\tz_2}-m_{\tz_1}$. By using suitable cuts,
it is possible to obtain hemispheric separation between the decay
products of the two gluinos in the $\eslt$ sample from
gluino pair production: the mass of the hadronic system in each hemisphere
measures~\cite{ETM} $m_{\tg}$ to 15-25\% provided $m_{\tg}\alt 700$~GeV.
A similar strategy had been suggested earlier~\cite{BGH} for same sign dilepton
events from gluino pair production, but just one production and
decay chain was examined, and that, at the parton level.

{\it Testing SUGRA Assumptions:} This is more difficult at hadron colliders.
However, because the model is specified by just four parameters plus a sign,
various signals become correlated. Since most sparticles should be accessible
at the LHC, it would be interesting to see if the various signals there
(as well as any signals in LEP2 or Tevatron experiments) can be accounted
for by a single set of parameters.

{\it Identifying Sparticle Sources:} Since several sparticles will
simultaneously be produced at the LHC, it is necessary to identify
observables or cuts to separate out these various production mechanisms.
Trileptons from $\tw_1\tz_2$ production~\cite{TRILEP} and
dileptons from slepton production~\cite{PREP}
can be isolated by suitable choices of cuts. It should also
be possible to tell
whether or not squarks
are being produced along with gluinos.
The presence of squarks in significant numbers will be signalled
by a charge asymmetry ($n(\ell^+\ell^+) > n(\ell^-\ell^-)$)
in the SS dilepton sample~\cite{BTW,ATLAS}, and also by a somewhat lower
jet multiplicity (compared to expectations for the measured gluino mass)
in the $\eslt$ sample~\cite{ETM} than expected from
just gluino production.
An excess of like-flavour $\ell^+ \ell^-$ pairs
as compared to $e^{\pm}\mu^{\mp}$ pairs would signal~\cite{BDKNT} a significant
production of neutralinos in cascade decays.
Finally, for some ranges of parameters, it is
also possible~\cite{ETM} to search for Higgs bosons from cascade
decays of gluinos via a bump in the $m_{b\bar{b}}$ distribution
in the SUSY-enriched $\eslt$ sample.

\section{Acknowledgments}
Collaboration with H.~Baer, M.~Bisset, M.~Brhlik, C-H.~Chen, M.~Drees, J.~Feng,
C.~Kao, R.~Munroe, H.~Murayama, M.~Nojiri,
F.~Paige, M.~Peskin, J.~Sender and J.~Woodside
as well as discussion with them and many other colleagues are gratefully
acknowledged. This research was supported in part by the U.S. Department
of Energy grant DE-FG-03-94ER40833.

\setcounter{secnumdepth}{0}

\end{document}